\documentclass[pre,twocolumn]{revtex4-1}

\usepackage{setspace}
\usepackage{color}
\usepackage{epsf}
\usepackage{graphicx}
\usepackage{amssymb}
\usepackage{amsbsy}
\usepackage{mathtools}

\newcommand{\revis}[1]{\textcolor{black}{#1}}

\DeclareFontFamily{OT1}{pzc}{}
\DeclareFontShape{OT1}{pzc}{m}{it}%
              {<-> s * [1.20] pzcmi7t}{}
\DeclareMathAlphabet{\mathpzc}{OT1}{pzc}%
                                 {m}{it}

\newcommand*\diff{\mathop{}\!\mathrm{d}}

\begin{document}
\title{Surveying an Energy Landscape}
\author{Stefan Schnabel}
\email{stefan.schnabel@itp.uni-leipzig.de}
\author{Wolfhard Janke}
\email{wolfhard.janke@itp.uni-leipzig.de}
\affiliation{Institut f\"ur Theoretische Physik, Universit\"at Leipzig, IPF 231101, 04081 Leipzig, Germany}
\date{\today}
\begin{abstract}
We derive a formula that expresses the density of states of a system with continuous degrees of freedom as a function of microcanonical averages of squared gradient and Laplacian of the Hamiltonian. This result is then used to propose a novel flat-histogram Monte Carlo algorithm, which is tested on a \revis{three-dimensional} system of interacting Lennard-Jones particles, the O($n$) vector spin model \revis{on hypercubic lattices in $D=1$ to $5$ dimensions, and the O($3$) Heisenberg model on a triangular lattice featuring frustration effects.}
\end{abstract}
%
\maketitle
\section{Introduction}

Central to statistic physics are the canonical ensemble, its partition function, and as defining quantity the temperature on the one hand as well as the microcanonical ensemble and the density of states defined by energy on the other. While the energy in the canonical ensemble takes the form of a weighted average over all microstates, temperature in the microcanonical ensemble becomes the inverse of the derivative of the logarithmic density of states. However, in 1997 Rugh \cite{Rugh} showed that it too can be expressed as an average of an observable comprising various derivatives of the Hamiltonian. The same formula had already been found \cite{dos_deriv}, but -- judging by the number of citations -- received very little attention. With Rugh's work at hand it was soon realized \cite{MC_test} that this relation among other things offers a way to verify the correctness of Monte Carlo (MC) computer simulations. Unfortunately, the formula involves a rather unwieldy term containing the Hamiltonian's Hessian that is computationally demanding. In this study, we develop a formula that expresses the density of states of the system as a function of microcanonical averages of squared gradient and Laplacian of the Hamiltonian while avoiding this term.

In the last decades MC simulations have become an important tool to investigate thermodynamic properties of models of complex systems. Today, many different techniques are used. In addition to the famous Metropolis algorithm \cite{metropolis} which samples from the canonical ensemble nowadays generalized ensemble methods have become more prevalent. Among these, flat-histogram method such as multicanonical  (MUCA) \cite{muca}, the Wang-Landau method \cite{wanglandau}, and Statistical Temperature MC \cite{STMC} aim to create the same ensemble where the distribution as a function of energy is constant. On the one hand this allows to reweight the data to a canonical ensemble with any desirable temperature, while on the other hand even if one is only interested in low-temperature behavior the inclusion of high energies allows the system to decorrelate more easily. To bias the system's random walk such that this ensemble becomes the stationary distribution the logarithm of the density of states (DoS) or its derivative must be known with sufficient precision. This can be achieved in an iterative process \cite{muca_it,multihistrew} that analyses and incorporates successively created histograms or on the fly by permanently altering the estimate of the DoS \cite{wanglandau} or its derivative \cite{STMC} based only on the energy of the current state of the system. Either way, the estimate of the DoS that is created and employed to drive the algorithm solely based on the energy time series. No other information is utilized. 

It is an interesting exercise and test of the accuracy and precision achieved with our formula to measure the microcanonical properties of the energy landscape and use them to estimate the DoS during a flat-histogram MC simulation that at the same time is using that estimate to achieve a flat distribution in energy. In this study we demonstrate this idea with two examples: A system of interacting Lennard-Jones particles and the O($n$) vector spin model.

The paper is organized as follows: In section 2 we once more derive the formula of Gilat and Rugh and proceed to calculate our alternative. In section 3 we discuss how a flat-histogram algorithm can be formulated and develop a suitable method for numerical integration. We then apply the method to a Lennard-Jones system in section 4 and to the O($n$) vector spin model in section 5. In section 6 we finish with some concluding remarks.

\section{Calculating the density of states}
The DoS or partition sum of the microcanonical ensemble at energy $E$ is given by
\begin{equation}
	g(E) = \int \delta(\mathpzc{H}(\mathbf{X})-E)\diff x^N,
\end{equation}
where $\mathpzc{H}$ is the Hamiltonian of the system, $N$ the number of degrees of freedom, and the integration goes over the entire state space. This can be rewritten as a surface integral
\begin{equation}
	g(E) = \int_{A_E} \frac{1}{|\nabla\mathpzc{H}(\mathbf{X})|}\diff x^{N-1},
\label{eq:eq_A}
\end{equation}
with $A_E = \{\mathbf{X} : \mathpzc{H}(\mathbf{X})=E\}$ being the surface of constant energy $E$ and $\nabla$ is the gradient $(\frac{\diff}{\diff x_1},\frac{\diff}{\diff x_2},...,\frac{\diff}{\diff x_N})^T$. The microcanonical average of any observable $Q(\mathbf{X})$ is given by
\begin{eqnarray}
	\langle Q\rangle_E =\frac{1}{g(E)}\int\delta(\mathpzc{H}(\mathbf{X})-E)Q(\mathbf{X})\diff x^N\\
        = \frac{1}{g(E)}\int_{A_E} \frac{Q(\mathbf{X})}{|\nabla\mathpzc{H}(\mathbf{X})|}\diff x^{N-1}.
\end{eqnarray}
To be able to apply the divergence theorem we rewrite (\ref{eq:eq_A}) again:
\begin{equation}
	g(E) = \int_{A_E} \frac{\nabla\mathpzc{H}(\mathbf{X})}{(\nabla\mathpzc{H}(\mathbf{X}))^2}\cdot\mathbf{n}(\mathbf{X})\diff x^{N-1}.
        \label{eq_dos_srf}
\end{equation}
Here, $\mathbf{n}=\nabla\mathpzc{H}(\mathbf{X})/|\nabla\mathpzc{H}(\mathbf{X})|$ is a unit vector perpendicular to the plane of constant energy and pointing to higher energies. It is therefore parallel to $\nabla\mathpzc{H}(\mathbf{X})$.
The derivative of the DoS with respect to energy is
\begin{equation}
	\frac{\diff g(E)}{\diff E} = \lim_{\varepsilon\rightarrow 0}\frac{g(E+\varepsilon)-g(E)}{\varepsilon}.
\end{equation}
We insert (\ref{eq_dos_srf}), apply the divergence theorem and integrate perpendicular to the surfaces by multiplying the thickness of the integration volume $\frac{\varepsilon}{|\nabla\mathpzc{H}(\mathbf{X})|}$ and find
\begin{equation}
	\frac{\diff g(E)}{\diff E} = \int_{A_E} \frac{1}{|\nabla\mathpzc{H}(\mathbf{X})|}\nabla\frac{\nabla\mathpzc{H}(\mathbf{X})}{(\nabla\mathpzc{H}(\mathbf{X}))^2}\diff x^{N-1}.
\end{equation}
Dividing by $g(E)$ on both sides we obtain the known result \cite{dos_deriv,Rugh,Giardina,Nurdin,Nurdin2,Gutierrez}
\begin{equation}
	\frac1{g(E)} \frac{\diff g(E)}{\diff E} = \frac{\diff\ln g(E)}{\diff E} = \left\langle B(\mathbf{X})\right\rangle_{E}
\end{equation}
with
\begin{eqnarray}
B(\mathbf{X})&=&\nabla\frac{\nabla\mathpzc{H}(\mathbf{X})}{(\nabla\mathpzc{H}(\mathbf{X}))^2}\nonumber \\
&=&\frac{\Delta\mathpzc{H}(\mathbf{X})}{(\nabla\mathpzc{H}(\mathbf{X}))^2}-2\frac{\nabla\mathpzc{H}(\mathbf{X})H(\mathbf{X})\nabla\mathpzc{H}(\mathbf{X})}{(\nabla\mathpzc{H}(\mathbf{X}))^4},
\end{eqnarray}
where $\Delta=\sum_{i=1}^N\frac{\partial^2}{\partial x_i^2}$ is the Laplace operator and $H$ denotes the Hessian matrix of the Hamiltonian, $H_{ij}(\mathbf{X})=\frac{\partial^2\mathpzc{H}(\mathbf{X})}{\partial x_i\partial x_j}$. The observable $B$ which can in principle be calculated for any microstate $\mathbf{X}$ of the system at hand, allows us to determine the DoS up to a factor regardless of the applied algorithm:
\begin{equation}
	g(E)\propto \exp\left( \int_{E_0}^E\langle B\rangle_{E'}\diff E'\right),
\label{eq:g_of_E_compl}
\end{equation}
where $E_0$ can be chosen freely. It is worth noting that $B$ relates directly to temperature. While it is true by definition that its microcanonical average equals the inverse microcanonical temperature if the latter is defined as
\begin{equation}
  (T_{\rm micro}k_{\rm B})^{-1} = \frac{\diff S_{\rm micro}}{\diff E},\ S_{\rm micro} =\ln g(E)
\end{equation}
it can also easily be shown that its canonical average is equal to the inverse canonical temperature:
\begin{eqnarray}
 \frac{\int B(\mathbf{X})e^{-\beta E(\mathbf{X})}\diff x^N }{ \int e^{-\beta E(\mathbf{X})}\diff x^N} &=& \frac{\int \langle B\rangle_{E'}g(E')e^{-\beta E'}\diff E' }{ \int g(E')e^{-\beta E'}\diff E'} \nonumber\\
                                      &=& \frac{\int g'(E')e^{-\beta E'}\diff E' }{ \int g(E')e^{-\beta E'}\diff E'} \nonumber\\
                                      &=& \beta = (k_{\rm B}T)^{-1}.
\end{eqnarray}

While gradient and Laplace operator can be applied to $\mathpzc{H}$ without too much computational effort\footnote{Here, we assume that the potentials are not uncommonly complicated.} the determination of the Hessian matrix is very demanding and a simpler scheme would be preferable. We start again with the microcanonical average of some observable $Q(\mathbf{X})$
\begin{equation}
  \langle Q(\mathbf{X})\rangle_E =\frac{1}{g(E)}\int_{A_E}\frac{Q(\mathbf{X})}{|\nabla\mathpzc{H}(\mathbf{X})|}\diff x^{N-1}\\
\end{equation}
and now consider its energy derivative
\begin{eqnarray}
  \frac{\diff}{\diff E}\langle Q(\mathbf{X})\rangle_E =\frac{\diff}{\diff E}\frac{\int_{A_E}\frac{Q(\mathbf{X})}{|\nabla\mathpzc{H}(\mathbf{X})|}\diff x^{n-1}}{g(E)}\nonumber\\
  =\frac{\frac{\diff}{\diff E}\int_{A_E}\frac{Q(\mathbf{X})}{|\nabla\mathpzc{H}(\mathbf{X})|}\diff x^{n-1}}{g(E)} - \langle Q(\mathbf{X})\rangle_E\frac{g'(E)}{g(E)}.
\end{eqnarray}
The integral can be transformed,
\begin{equation}
  \int_{A_E}\frac{Q(\mathbf{X})}{|\nabla\mathpzc{H}(\mathbf{X})|}\diff x^{n-1}=\int_{A_E}\frac{Q(\mathbf{X})}{(\nabla\mathpzc{H}(\mathbf{X}))^2}\nabla\mathpzc{H}(\mathbf{X})\cdot\mathbf{n}\diff x^{N-1},
\end{equation}
and the derivative be calculated similar to the procedure used above. Using differential quotient and divergence theorem we find
\begin{equation}
  \frac{\diff}{\diff E}\int_{A_E}\frac{Q(\mathbf{X})}{|\nabla\mathpzc{H}(\mathbf{X})|}\diff x^{n-1} = \int_{A_E}\frac{\nabla\left(\frac{Q(\mathbf{X})\nabla\mathpzc{H}(\mathbf{X})}{(\nabla\mathpzc{H}(\mathbf{X}))^2} \right)}{|\nabla\mathpzc{H}(\mathbf{X})|}\diff x^{N-1}.
\end{equation}
It follows that
\begin{equation}
\frac{\diff}{\diff E}\langle Q(\mathbf{X}) \rangle_E = \left\langle \nabla\left(\frac{Q(\mathbf{X})\nabla\mathpzc{H}(\mathbf{X})}{(\nabla\mathpzc{H}(\mathbf{X}))^2} \right) \right\rangle_E 
                                                     - \langle Q(\mathbf{X})\rangle_E\frac{\diff\ln g(E)}{\diff E}
\end{equation}
and in particular by choosing $Q(\mathbf{X})=(\nabla\mathpzc{H}(\mathbf{X}))^2$ one obtains
\begin{equation}
\frac{\diff}{\diff E}\langle (\nabla\mathpzc{H}(\mathbf{X}))^2\rangle_E = \langle \Delta\mathpzc{H}(\mathbf{X})\rangle_E - \langle (\nabla\mathpzc{H}(\mathbf{X}))^2\rangle_E\frac{d\ln g(E)}{\diff E}.
\end{equation}
We obtain for the inverse microcanonical temperature:
\begin{equation}
\frac{\diff \ln g(E)}{\diff E} = \frac{\langle \Delta\mathpzc{H}(\mathbf{X})\rangle_E}{\langle (\nabla\mathpzc{H}(\mathbf{X}))^2\rangle_E} - \frac{\diff}{\diff E}\ln\langle (\nabla\mathpzc{H}(\mathbf{X}))^2\rangle_E.
\end{equation}
Integrating on both sides and exponentiating gives
\begin{equation}
g(E) \propto \frac{1}{\langle(\nabla\mathpzc{H}(\mathbf{X}))^2\rangle_E}\exp\left(\int_{E_0}^E \frac{\langle \Delta\mathpzc{H}(\mathbf{X})\rangle_{E'}}{\langle (\nabla\mathpzc{H}(\mathbf{X}))^2\rangle_{E'}}\diff E'\right) .
\label{eq:eq:g_of_E_main}
\end{equation}
We first derived (\ref{eq:eq:g_of_E_main}) in a  different way: If one formally considers a random walk in configuration space with sufficiently small step length and equilibrates the system after every single step on the respective surface of constant energy, one obtains a one-dimensional stochastic process in energy with a stationary distribution proportional to $g(E)$. The change in energy for a small step $\mathbf X \rightarrow \mathbf X'=\mathbf X+\mathbf x$ is given by
\begin{equation}
E'-E=\mathbf x\nabla \mathpzc{H}(\mathbf{X})+\frac12 \mathbf x H(\mathbf{X}) \mathbf x + O(|\mathbf x|^3)
\label{eq:E_change}
\end{equation}
and it follows that the drift for such a process is $\frac{\alpha}{2}\langle \Delta\mathpzc{H}(\mathbf{X})\rangle_{E}$ and the diffusion $\alpha\langle (\nabla\mathpzc{H}(\mathbf{X}))^2\rangle_{E}$, where $\alpha$ is a constant related to the length of $\mathbf x$. In this context (\ref{eq:eq:g_of_E_main}) represents the solution of the Fokker-Planck equation.

In the context of MC simulations the DoS is virtually always determined via histograms. The distribution of energies within the employed ensemble is estimated and allows to calculate $g(E)$. Although rare, faulty simulations with the detailed balance criterion violated are not unheard-of and it is sometimes not easy to spot such problems. It is worth noting that (\ref{eq:g_of_E_compl}) and (\ref{eq:eq:g_of_E_main}) provide an alternative way to determine the DoS and a comparison with the histogram-derived DoS can, therefore, be used to test whether an algorithm is in balance.

\section{Algorithm}
It is well known and widely used that within a Monte Carlo simulation a flat histogram can be produced if the acceptance probability for proposed moves is given by
\begin{equation}
P_{\rm acc}(E_{\rm old},E_{\rm new})=\min\left(1,\frac{g(E_{\rm old})}{g(E_{\rm new})}\right),
\end{equation}
which can now be written as
\begin{equation}
P_{\rm acc}(E_{\rm old},E_{\rm new})=\min\Bigg(1,\frac{\langle(\nabla\mathpzc{H})^2\rangle_{E_{\rm new}}}{\langle(\nabla\mathpzc{H})^2\rangle_{E_{\rm old}}} \exp\int_{E_{\rm new}}^{E_{\rm old}}\frac{\langle \Delta\mathpzc{H} \rangle_{E'}}{\langle (\nabla\mathpzc{H})^2\rangle_{E'}}\diff E'\Bigg).
\label{eq:pacc}
\end{equation}
The arguments $\mathbf{X}$ have been removed for the sake of clarity.

One main challenge is the accurate numeric evaluation of the integral. Since the energy is continuous, it is natural to employ a binning procedure. It might be worthwhile to use an adaptive binwidth with higher resolution in regions where the integrand 
\begin{equation}
f(E)=\frac{\langle \Delta\mathpzc{H} \rangle_{E}}{\langle (\nabla\mathpzc{H})^2\rangle_{E}}
\end{equation}
changes rapidly with $E$, but here we use intervals $I_k=[E_0+(k-1)h,E_0+kh]$ of constant width $h$ and estimate microcanonical averages of an observable $O$ as the mean of all measurements with an energy $E_t\in I_k$:
\begin{equation}
\langle O\rangle_E \approx [O]_k \coloneqq \frac{ \sum_{E_t\in I_k} O_t }{ \sum_{E_t\in I_k} 1 },
\label{eq:O_av}
\end{equation}
where $t$ is the time index of the measurements. It is useful to also measure $[E]_k$ and use it instead of the midpoint of $I_k$ for the integration. We use the notation $E_k=[E]_k$ and $f_k=[ \Delta\mathpzc{H} ]_k/[ (\nabla\mathpzc{H})^2 ]_k$.

Following the standard approach for quadrature (numerical integration) we locally approximate the data by an analytical function and integrate the latter. However, the usual choice of polynomials does not represent the underlying mathematical relation well. Since it is 
\begin{equation}
f(E) \approx \frac{\diff}{\diff E}\ln g(E),
\label{eq:fg_approxim}
\end{equation}
we use the Ansatz
\begin{equation}
g(E) \propto |E-\eta|^\mu,
\end{equation}
which corresponds to a system with lowest energy $\eta$ and constant (canonical) specific heat $C=k_{\rm B}(\mu+1)$. Assuming equality in (\ref{eq:fg_approxim}) it is
\begin{equation}
f(E)=\frac{\mu}{E-\eta}
\end{equation}
and for two points $(E_i,f_i)$ and $(E_{i+1},f_{i+1})$ we obtain
\begin{eqnarray}
\mu_i &=& \frac{E_i-E_{i+1}}{f_i^{-1}-f_{i+1}^{-1}},\\
\eta_i  &=& \frac{E_if_i-E_{i+1}f_{i+1}}{f_i-f_{i+1}}.
\end{eqnarray}
Thus we arrive at
\begin{equation}
\int_{E_k}^{E_l}f(E)dE\approx\sum_{i=k}^{l-1}\mu_i\ln\left|\frac{E_{i+1}-\eta_i}{E_{i}-\eta_i}\right|.
\end{equation}
For the actual simulation we use the function
\begin{equation}
G(E) = \int_{E_0}^{E}\frac{\langle \Delta\mathpzc{H} \rangle_{E'}}{\langle (\nabla\mathpzc{H})^2\rangle_{E'}}\diff E'
\end{equation}
with some suitable $E_0$ and as shown above calculate $G(E_k)$ for all bins (intervals) $I_k$.
Inside each bin we approximate \revis{$G(E) \approx G(E_k)+G'(E_k)(E-E_k) $} linearly by using $G(E_k)$ from the numerical integration and $G'(E_k)=f_k$.
\revis{The acceptance probability from} eq.~(\ref{eq:pacc}) becomes
\begin{equation}
P_{\rm acc}(E_{\rm old},E_{\rm new})=\min\Bigg(1, \frac{\langle(\nabla\mathpzc{H})^2\rangle_{E_{\rm new}}  }{ \langle(\nabla\mathpzc{H})^2\rangle_{E_{\rm old}} } \exp\left[G(E_{\rm old})-G(E_{\rm new})\right] \Bigg).
\end{equation}
We simulate and measure $\nabla\mathpzc{H}$ and $\Delta\mathpzc{H}$ for a short while using $G(E)$, recalculate it, and repeat. Of course, this introduces small violations of the detailed balance criterion, albeit to a much lesser extent than a Wang-Landau simulation. Nevertheless, as with any flat histogram simulation the final data should be produced in a simulation with fixed $G(E)$ and $\langle(\nabla\mathpzc{H})^2\rangle_{E}$.

As we will show in the next section, in the form presented the algorithm is able to simulate quite large systems. However, there also is a drawback. The estimate of the DoS is necessarily based on information that has been gathered only in regions of state space that already have been sampled. This can include states that represent rare events in the converged ensemble, e.g., configurations that correspond to a supercritical gas. If the ``correct'' state -- the condensate or droplet in the example -- hasn't been found yet, then the estimates of microcanonical averages are dominated by the ``wrong'' data and it can take a very long time to correct this bias. Thus first-order phase transitions or rough energy landscapes can pose a challenge for the algorithm in its basic form. A more refined method of averaging than eq. (\ref{eq:O_av}) which  attributes higher weight to later measurements might turn out to be a solution for this problem.

\section{Lennard-Jones particles}

Clusters of Lennard-Jones particles and their morphology at low temperature have been under study for a long time. Modeling noble gas atoms, Lennard-Jones particles are an interesting object of inquiry in their own right and they provide challenging benchmark systems for numerical optimization since their energy landscape contains numerous minima belonging to competing geometric structures. For small sizes the ground states have been determined some time ago \cite{LJC_GS_147,LJC_GS_Wales,LJC_GS} and the behavior is well understood. If the number of atoms is a few thousands or less, the low-temperature phase is dominated by icosahedral geometry \cite{ico}. In many cases there are solid-solid transitions \cite{LJC_MM_trans} where the outer layer of the cluster changes from a so-called anti-Mackay shape that maximizes entropy to a Mackay structure minimizing energy. In some rare cases $N=38,75,76,77,98,102,103,104,\dots$ non-icosahedral states are occupied at a very low temperature leading to solid-solid transitions that can be extremely challenging to investigate by means of MC simulations \cite{LJ_poly_method}.

We consider $N$ particles in three dimensional-space which interact pairwise through a 12-6 Lennard-Jones potential
\begin{equation}
U(r)=\frac1{r^{12}}-\frac2{r^6}.
\end{equation}
With this parametrization the potential has its minimum at $r_0=1$.    
The particles are freely mobile within a cubic volume of linear extension $L$ and we label their positions as $\mathbf x\in [0,L]^3$. The Hamiltonian reads
\begin{equation}
\mathcal{H}=\sum_{i=1}^{N-1}\sum_{j=i+1}^{N}U(|\mathbf x_i-\mathbf x_j|).
\end{equation}
One finds that
\begin{equation}
\nabla_i\mathcal{H}=-12\sum_{j\ne i}(\mathbf x_i-\mathbf x_j)\left( \frac1{|\mathbf x_i-\mathbf x_j|^{14}}-\frac1{|\mathbf x_i-\mathbf x_j|^8} \right)
\end{equation}
and calculating or refreshing
\begin{equation}
(\nabla\mathcal{H})^2=\left(\sum_{i=1}^N \nabla_i\mathcal{H}\right)^2
\end{equation}
is, therefore, somewhat cumbersome. Thankfully,
\begin{equation}
\Delta\mathcal{H}=\sum_{i=1}^{N-1}\sum_{j=i+1}^{N}24\left( \frac{11}{|\mathbf x_i-\mathbf x_j|^{14}}-\frac5{|\mathbf x_i-\mathbf x_j|^8} \right)
\end{equation}
is simpler.

\begin{figure}
\begin{center}
\includegraphics[width=0.95\columnwidth]{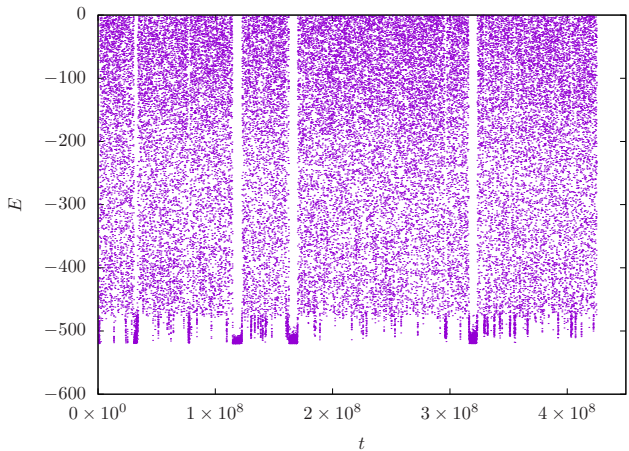}
\caption{{\label{fig:ts_100csl} Time series of the energy $E$ throughout a simulation of $N=100$ Lennard-Jones particles. The energy was restricted to $E>-520$.}}
\end{center}
\end{figure}

\begin{figure}
\begin{center}
\includegraphics[width=0.95\columnwidth]{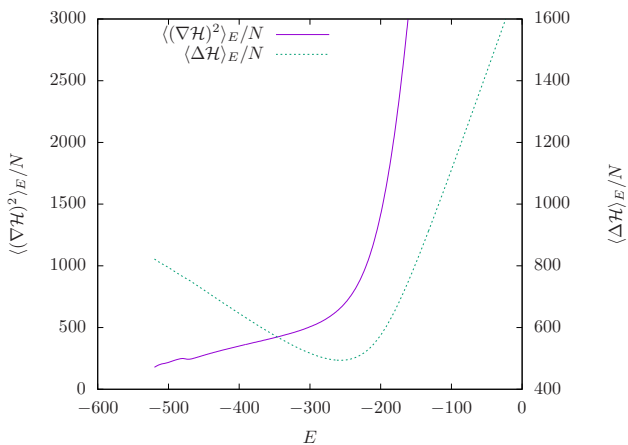}
\caption{{\label{fig:dd_100csl} The microcanonical averages $\langle(\nabla\mathpzc{H})^2\rangle_E/N$ and $\langle \Delta \mathpzc H \rangle_E/N$ from the same simulation as the data in Fig.~\ref{fig:ts_100csl}.}}
\end{center}
\end{figure}

\begin{figure}
\begin{center}
\includegraphics[width=0.95\columnwidth]{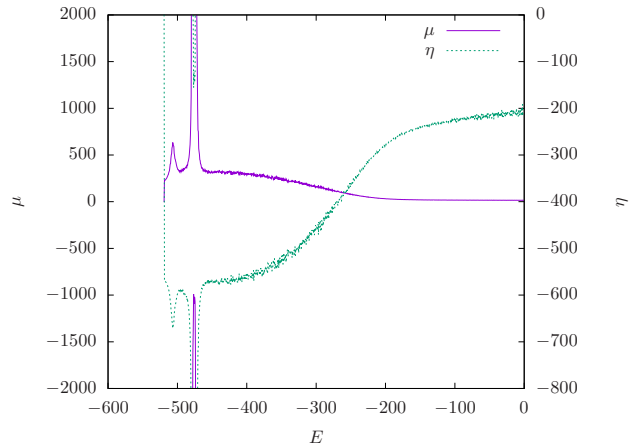}
\caption{{\label{fig:ip_100csl} The integration parameters $\mu$ and $\eta$ from the same simulation as the data in Fig.~\ref{fig:ts_100csl}.}}
\end{center}
\end{figure}

We performed a simulation of $N=100$ particles confined in a steric cube with $L=5r_0$. The ground-state energy of this system is $E_g=-557.039820$ \cite{LJC_GS_147} and we restrict the energy to $-520<E<0$. The energy as a function of simulation time in units of $N$ single-atom displacement moves can be seen in Fig.~\ref{fig:ts_100csl}. It is apparent that the simulation is able to reach all energies in the interval within a relatively short time. The early wedge-shaped blocks at low energy indicate that the averages are not converged yet and balance is established by repeatedly transitioning in and out of the low-energy state.
Fig.~\ref{fig:dd_100csl} shows the microcanonical averages $\langle(\nabla\mathpzc{H})^2\rangle_E/N$ and $\langle \Delta \mathpzc H \rangle_E/N$. Interestingly, the Laplacian shows a close to linear behavior throughout most of the energy interval, while squared gradient, as one would expect, goes to zero as $E$ approaches the ground-state energy. Its graph also displays an inflection, signaling a transition.
The integration parameters $\mu$ and $\eta$ shown in Fig.~\ref{fig:ip_100csl} also strongly relate to the thermodynamic behavior of the system and might be used similarly to a microcanonical analyses of the density of states \cite{MicroCanAnalysis}. Since $\mu$ is closely related to the specific heat it behaves similarly. Kinetic degrees of freedom are not taken into account in the simulation and as a consequence we observe $\mu\approx0$ for high energies in the gas phase. The condensation transition towards a liquid droplet with a non-zero $\mu$ is rather weak due to the small system size. Around $E\approx-475$, $G(E)$ becomes concave which manifests as $\mu<0$. This signal indicates the first-order-like freezing transition which leads to the formation of an icosahedral structure \cite{ico}. We suspect that the remaining signal at $E\approx-507$ is caused by the rearrangement of surface atoms from a so-called anti-Mackay to a Mackay structure \cite{LJC_MM_trans}. All transitions also manifest in $\eta$. While $\eta(E)<E$ in most cases if $\mu<0$ then the local approximation of $G(E)$ does not become zero at $E=\eta$, but instead diverges. Since its slope is positive in these cases it is $\eta>E$.

\section{O(\MakeLowercase{$\pmb{n}$}) spin model}

The O($n$) spin model is the generalization of the Ising ($n=1$), XY ($n=2$), and Heisenberg ($n=3$) spin models. In this model spins $\pmb{\sigma}\in\mathbb{R}^n, |\pmb{\sigma}|=1$ are elements of the $n$-sphere and are positioned on sites of regular lattices and interact through the Hamiltonian
\begin{equation}
\mathcal{H}=-J\sum_{\langle ij\rangle} \pmb{\sigma}_i\pmb{\sigma}_j, 
\end{equation}
where the sum runs over all lattice bonds and $J$ is the spin-spin interaction strength.  
To evaluate $\nabla\mathpzc{H}$ and $\Delta\mathpzc{H}$ we first consider the contribution to the total energy by an individual spin $\pmb{\sigma}_k$:
\begin{equation}
e_k=-\pmb{\sigma}_k\mathbf h_k/|\pmb{\sigma}_k|,
\label{eq:spin_energy}
\end{equation}
where $\mathbf h_k$ is the local field
\begin{equation}
\mathbf h_k = J\sum_{j\in nb(k)}\pmb{\sigma}_j
\end{equation}
and $nb(k)$ the set of neighbors of spin $\pmb{\sigma}_k$. In the following we set $J=1$ and refrain from displaying it in the formulae. This is equivalent to assuming that $e_k, h_k$, and $E$ are measured in units of $J$.  It is convenient to divide by $|\pmb{\sigma}_k|$ in eq.~(\ref{eq:spin_energy}) and for the moment to \revis{relax the $n$-sphere constraint to} $|\pmb{\sigma}_k|\ne1$ since this allows one to use the one-particle gradient $\nabla_k=(\frac\partial{\partial\pmb{\sigma}_{k,1}},\dots,\frac\partial{\partial\pmb{\sigma}_{k,n}})^T$ in Cartesian coordinates with the radial component $\pmb{\sigma}_k(\nabla_ke_k)$ ensured to be zero. We find that
\begin{equation}
\nabla_ke_k|_{|\pmb{\sigma}_k|=1} = \mathbf h_k-(\mathbf h_k\pmb{\sigma}_k)\pmb{\sigma}_k,
\end{equation}
and since $\mathbf h_k-(\mathbf h_k\pmb{\sigma}_k)\pmb{\sigma}_k$, $(\mathbf h_k\pmb{\sigma}_k)\pmb{\sigma}_k$, and $\mathbf h_k$ form a right-angled triangle it follows \footnote{In the case of $n=3$ an alternative expression is $\mathbf h_k^2-(\mathbf h_k\pmb{\sigma}_k)^2=(\mathbf h_k\times \pmb{\sigma}_k)^2$ \cite{Nurdin}.}
\begin{eqnarray}
(\nabla_ke_k|_{|\pmb{\sigma}_k|=1})^2 &=& \mathbf h_k^2-(\mathbf h_k\pmb{\sigma}_k)^2,\\
                                       &=& \mathbf h_k^2-e_k^2.
\end{eqnarray}
If the system is homogeneous, i.e., if all spins are equivalent we can drop the index $k$ and if the total number of spins is given by $N$ it is 
\begin{equation}
\langle(\nabla\mathpzc H)^2\rangle_E = N( \langle\mathbf h^2\rangle_E - \langle e^2\rangle_E ).
\end{equation}
Next we calculate that
\begin{equation}
\Delta_ke_k|_{|\pmb{\sigma}_k|=1} = \nabla^2_ke_k|_{|\pmb{\sigma}_k|=1} = -(n-1)e_k
\end{equation}
and noting that $\langle e \rangle_E=2E/N$ we find
\begin{eqnarray}
\langle\Delta\mathpzc H\rangle_E &=& -N(n-1)\langle e \rangle_E,\\
                               &=& -2(n-1) E.
\end{eqnarray}
Finally we arrive at the surprisingly simple result
\begin{equation}
g(E) \propto \frac{1}{ \langle\mathbf h^2\rangle_E - \langle e^2\rangle_E }\exp{\left(\int_{E_0}^E \frac{-2(n-1) E'/N}{ \langle\mathbf h^2\rangle_{E'} - \langle e^2\rangle_{E'} }\diff E'\right)} .
\label{eq:g_of_E_On}
\end{equation}
Unfortunately, this formula does not generalize to the Ising model $n=1$ since on the one hand a continuous energy scale is implicitly required and on the other hand for the Ising model $\langle\mathbf h^2\rangle_E = \langle e^2\rangle_E$.

\begin{figure}
\begin{center}
\includegraphics[width=0.95\columnwidth]{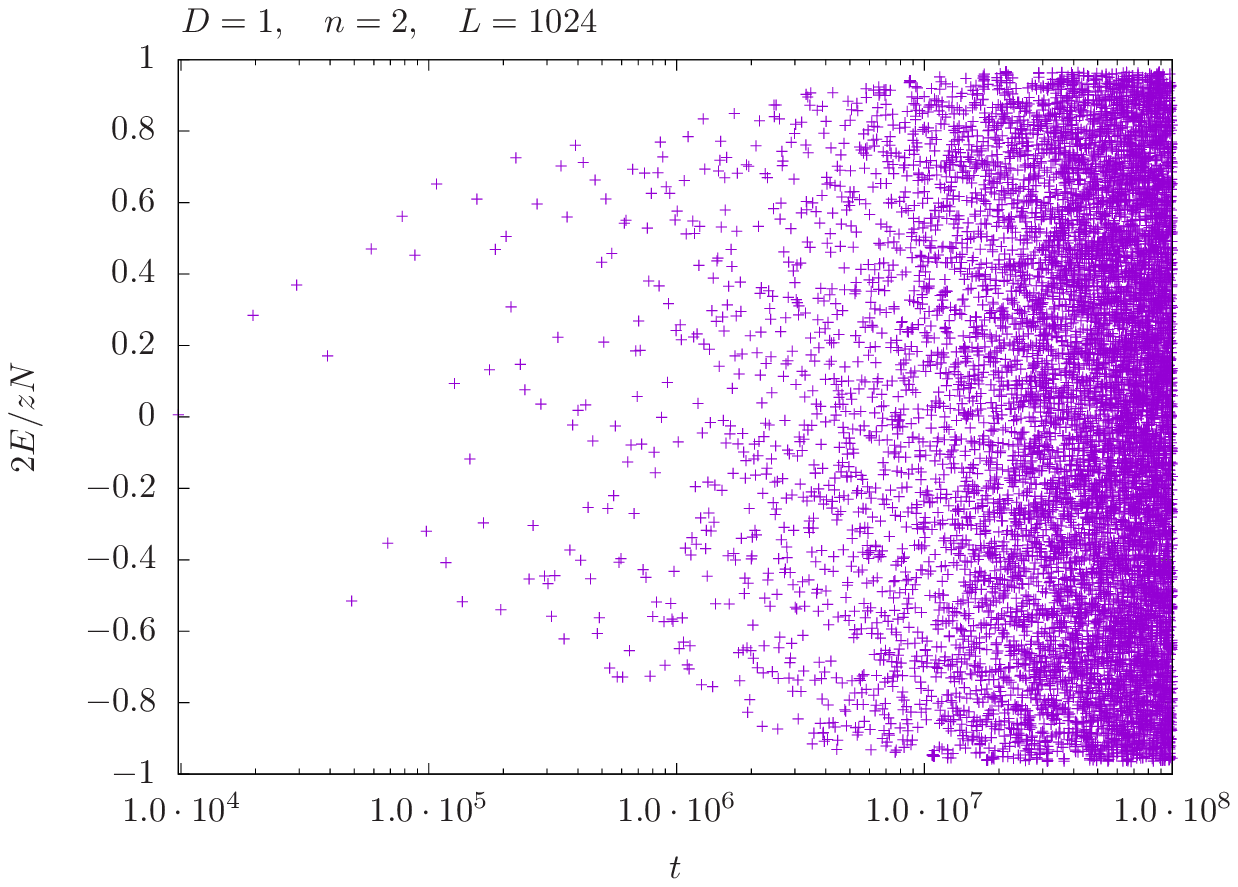}\\
\includegraphics[width=0.95\columnwidth]{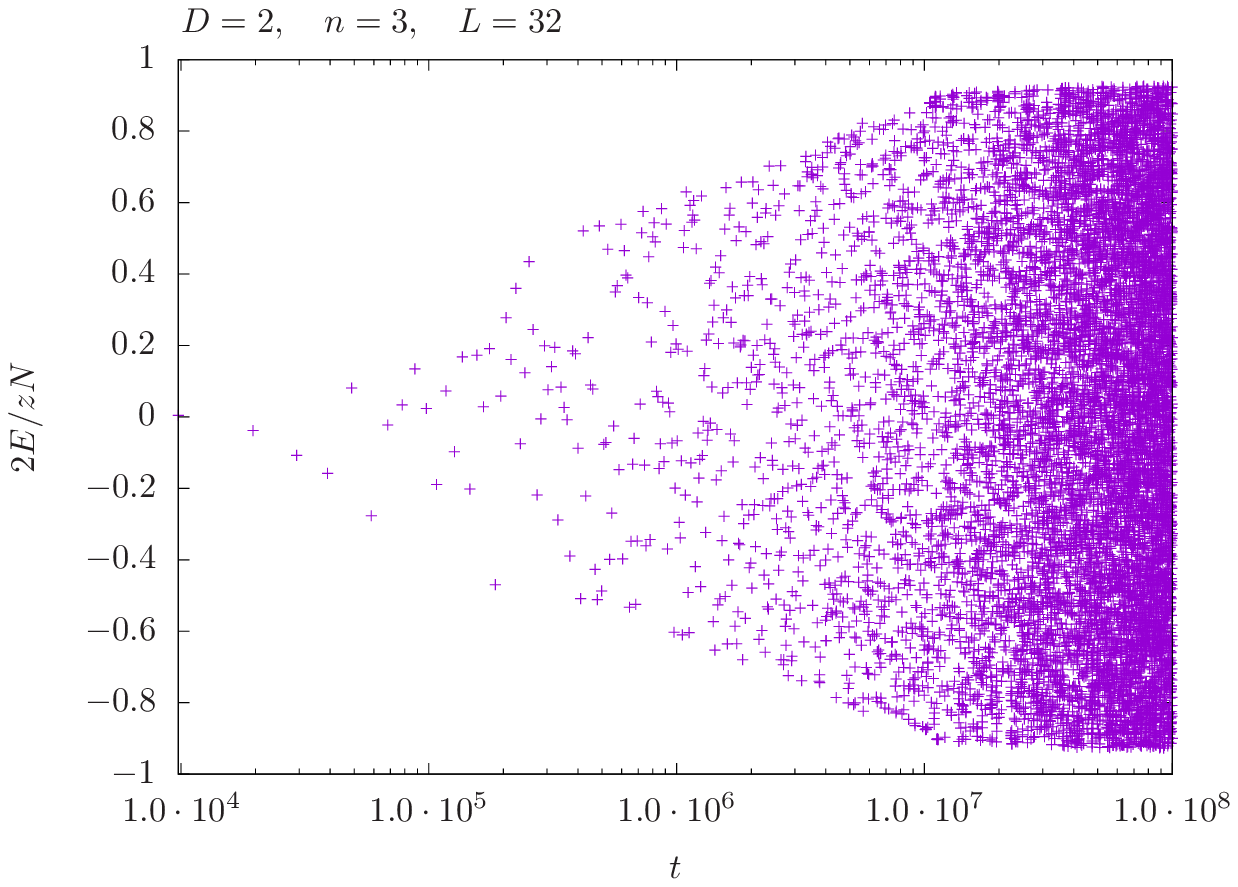}\\
\includegraphics[width=0.95\columnwidth]{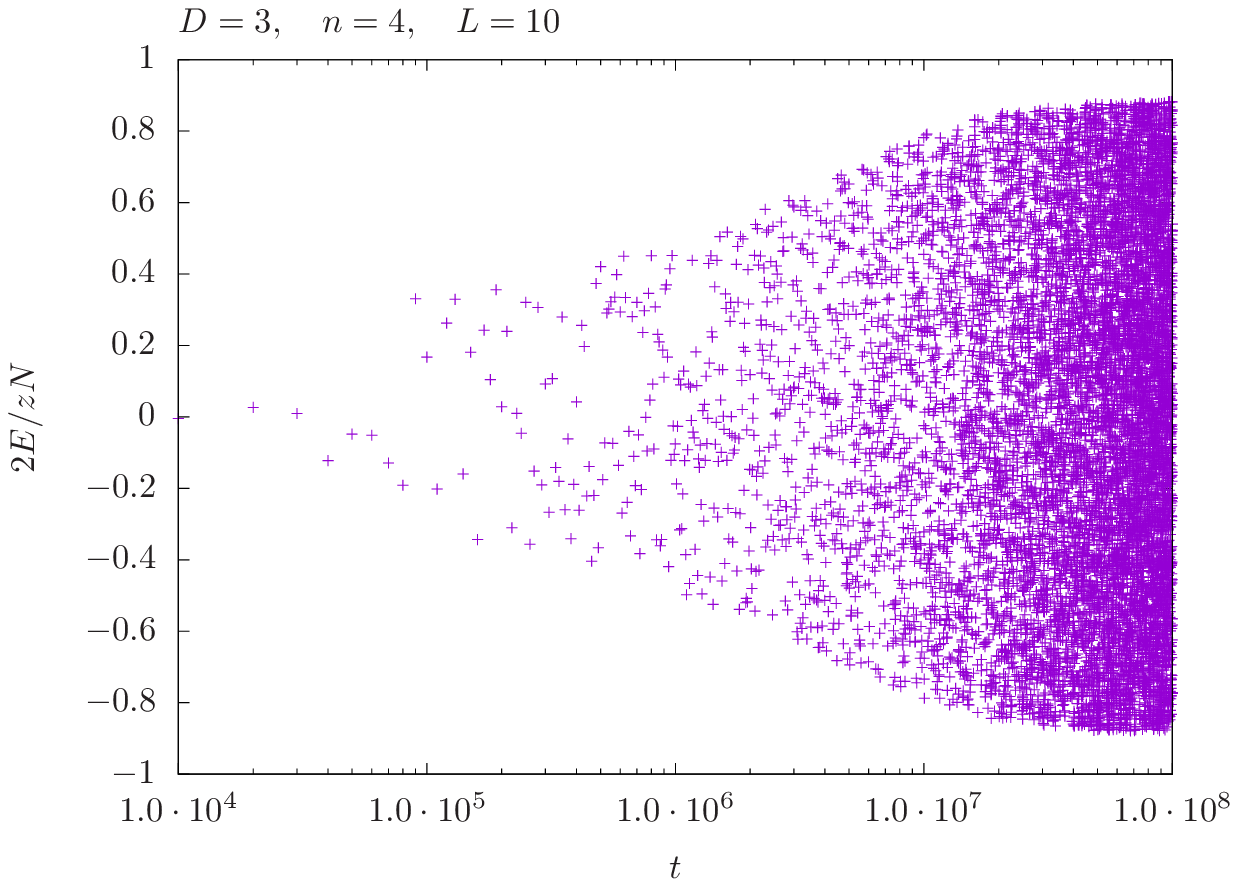}
\caption{{\label{fig:ts_On} Time series of the energy per bond $2E/zN$ throughout simulations of $N\approx1000$ spins for different values of $D$ and $n$. The time $t$ is measured in units of $N$ updates.}}
\end{center}
\end{figure}

\begin{figure}
\begin{center}
\includegraphics[width=0.95\columnwidth]{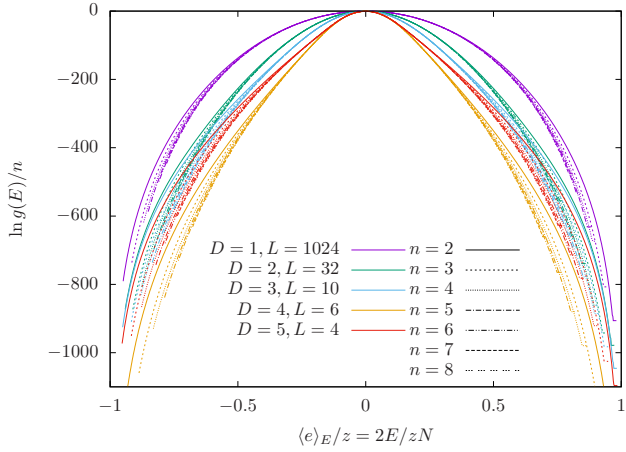}\\
\caption{{\label{fig:dos} Logarithmic density of states divided by $nN$ for different sets of values of $D, L,$ and $n$.}}
\end{center}
\end{figure}

We now consider hypercubic lattices in $D$ dimensions with linear extension $L$, $N=L^D$ spins, and periodic boundary conditions. For these lattices the number of neighbors of any site (spin), the so-called coordination number, is $z=2D$. During the simulation we use $N$ bins and integrate after every $10^3N$ individual spin updates. The single concern for selecting this value was to choose it large enough to not slow down the simulation by the computational effort of integrating. Proposed values for spins are selected randomly and independently of the current value. They are drawn using the rejection method for $n=2$, Marsaglia's methods \cite{Marsaglia} for $n=3,4$ and our own technique \cite{hypersphere} for $n>4$. Time series of the energy per bond $2E/zN$ for different values of $D$ and $n$ and about $N\approx10^3$ spins are shown in Fig.~\ref{fig:ts_On}.
For these cases the simulation is able to cover most of the available energy interval within about $10^7$ sweeps. We point out that for all simulations shown the ratio between the maximum of the DoS and the minimal value reached is between $10^{780}$ and $10^{2000}$. Of course such values can also be achieved with established state-of-the-art flat-histogram methods, but it is satisfying that this is possible with this method as well since it implies that the integration is done with adequate accuracy. The simulations fall a little bit short of the extremal energies $E_{\rm max}=-E_{\rm min}=ND$. We suspect that one reason is the comparatively large bin width which can become problematic if $G$ or its derivative become very steep. \revis{From the measured densities of states $g(E)\propto\exp[G(E)]/\langle (\nabla\mathpzc H)^2\rangle_E$ shown in Fig. \ref{fig:dos} it becomes apparent that due to the relatively low number of just 1000 bins, values in adjacent bins can differ by more than 20 orders of magnitude. It is satisfying that thanks to the linear interpolation of $G(E)$ a relatively flat distribution inside the bins can be maintained regardless and transitions between the bins are still occurring.} Another cause \revis{for decreasing performance at extremal energies} will certainly be the \revis{small} acceptance rate. Close to the minimal and maximal energy values spins are almost parallel to the local field and since we draw new spin values completely randomly the probability that such a proposal is accepted becomes very low. We refrained from optimizing the simulation since this study is mostly intended as a proof-of-concept. 

\begin{figure}
\begin{center}
\includegraphics[width=0.95\columnwidth]{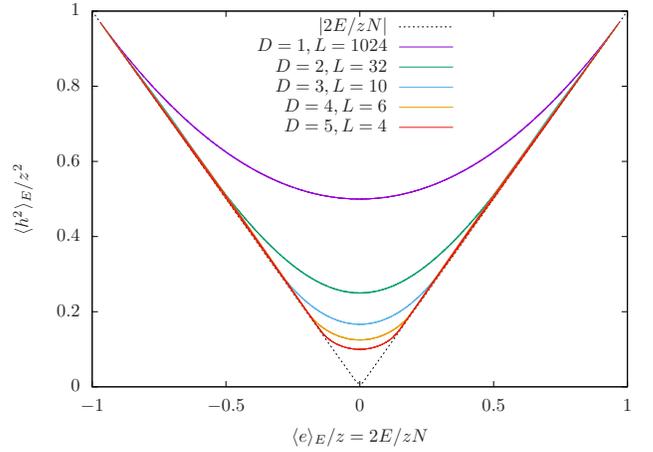}\\
\caption{{\label{fig:h2} Microcanonical average $\langle h^2\rangle_E$ as function of the energy per bond. For each $D$ curves are shown for $n=2,\dots,8$ which exhibit hardly any variation.}}
\end{center}
\end{figure}

We find that for large enough systems $\langle h^2\rangle_E$ depends only weakly on the dimension of spin space $n$. Note that the microcanonical ensemble in the thermodynamic limit directly fixes the correlation between neighboring spins
\begin{equation}
\langle\pmb{\sigma}_*\pmb{\sigma}_|\rangle_E = -\frac E{N}
\end{equation}
and $\langle h^2\rangle_E$ can be expressed in terms of correlations of next-nearest neighbor spins 
\begin{equation}
 \langle h^2\rangle_E=z+z\langle\pmb{\sigma}_*\pmb{\sigma}_{||}\rangle_E+z(z-2)\langle\pmb{\sigma}_*\pmb{\sigma}_{\llcorner}\rangle_E,
\end{equation}
where from any spin $\pmb{\sigma}_*$ the spins $\pmb{\sigma}_{|},\pmb{\sigma}_{||}$, and $\pmb{\sigma}_{\llcorner}$ are reached by one bond, two parallel bonds and two non-parallel bonds, respectively.
This allows one to show that in one dimension $\langle h^2\rangle_E$ is even independent of $n$ and one obtains 
\begin{equation}
\lim_{N\rightarrow\infty} \langle h^2 \rangle_E |_{D=1} = 2+2(E/N)^2
\label{eq:h2D1}
\end{equation}
which is in very good agreement with our data and would be indistinguishable from the graphs for $D=1$ in Fig.~\ref{fig:h2}. For the other values of $D$ all curves for different $n$ in Fig.~\ref{fig:h2} also are very close to identical. \revis{Separate simulations for $D=2,3$ at energies close to the transition revealed that in the thermodynamic limit the difference in $\langle h^2 \rangle_E$ for different values of $n$ is of the order of $1\%$.} This behavior is reminiscent of another case of unexpectedly small dependence on $n$: the critical energy density \cite{Nerattini}.

The situation is different for $\langle e^2\rangle_E$ which comprises $z$ second moments of nearest-neighbor spin products $\left\langle(\pmb{\sigma}_k\pmb{\sigma}_i)^2\right\rangle_E$ as well as $z(z-1)$ bond-bond correlations $\left\langle(\pmb{\sigma}_k\pmb{\sigma}_i)(\pmb{\sigma}_k\pmb{\sigma}_j)\right\rangle_E$. We are able to calculate the curves for $D=1$ and large $N$, but these do depend on $n$ (see Appendix). The data in Fig.~\ref{fig:e2}a suggest that for any $D$, large $n$ and $N$ an approximation may be given through
\begin{equation}
\langle e^2 \rangle_E \approx \left(\frac{2E}N\right)^2+\frac{zf_D(2E/zN)}{n}
\label{eq:e2}
\end{equation}
with additional corrections. Here, $f_D(x)$ is a function that can easily be calculated in $D=1$ dimension. We find
\begin{equation}
f_1(x)=\frac{(1-x^2)^2}{1+x^2}.
\label{eq:f1}
\end{equation}
However, it appears that this function is also valid for $D>1$ and we are led to believe by Fig.~\ref{fig:e2}b that the next correction is of the order $z/n^{1/D}$. This is of course a somewhat speculative heuristic analysis and even though the systems are of medium size $N\approx10^3$ the linear extension of the lattices for $D>2$ is small.
\begin{figure}
\begin{center}
\includegraphics[width=0.75\columnwidth]{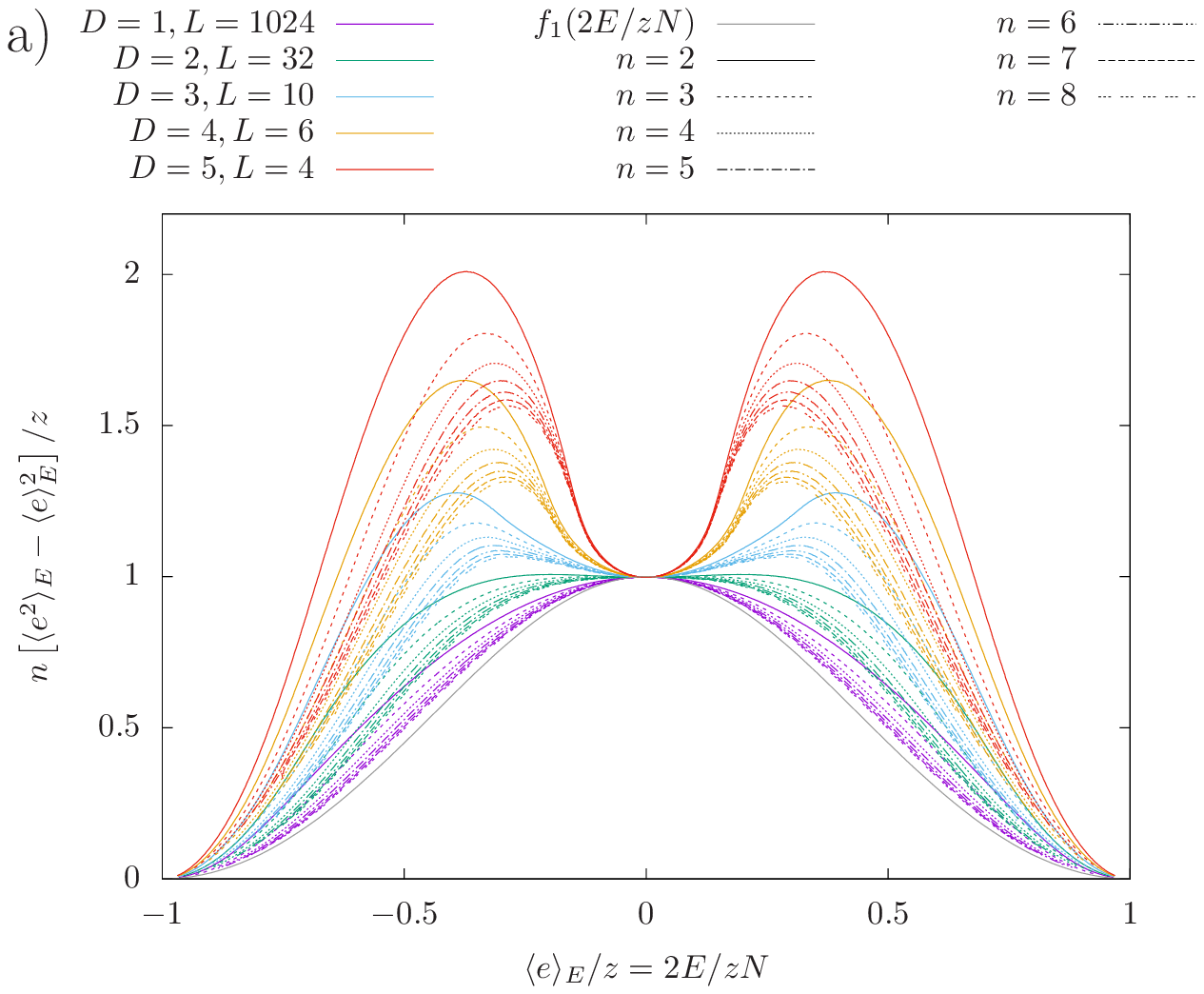}\\
\vspace{.3cm}
\includegraphics[width=0.75\columnwidth]{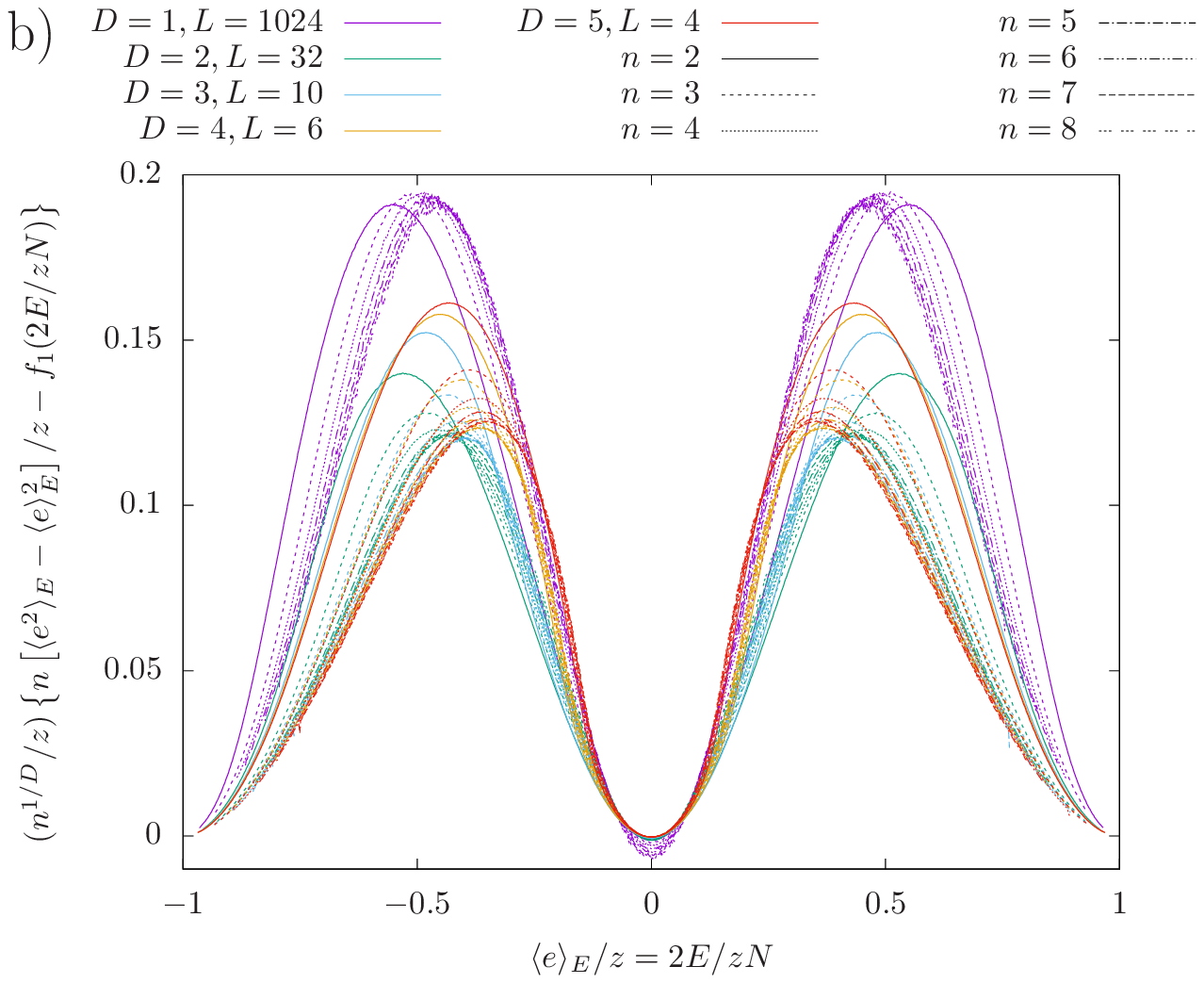}\\
\caption{{\label{fig:e2} (a) The difference of the measured microcanonical average of $\langle e^2\rangle_E$ and the squared spin energy multiplied by $n/z$ for different sets of values of $D, L, n$. The gray line represents the theoretical function $f_1$ for $D=1, n\rightarrow\infty$ and $L\rightarrow\infty$ given in Eq. (\ref{eq:f1}). (b) The difference between the data in (a) and $f_1$ appears to be approximately proportional to $z/n^{1/D}$.}}
\end{center}
\end{figure}

\begin{figure}
\begin{center}
\includegraphics[width=0.95\columnwidth]{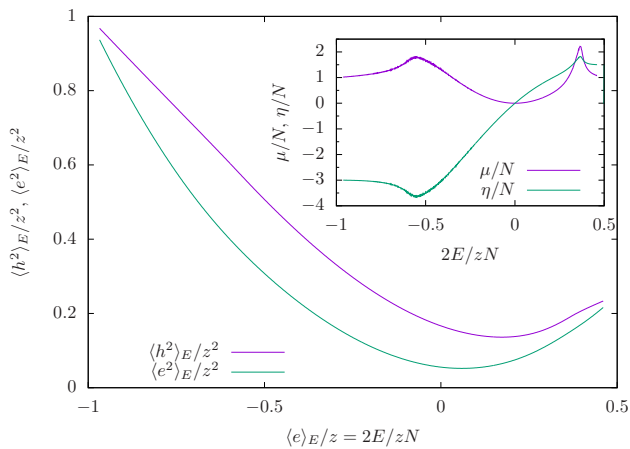}\\
\caption{{\label{fig:tri} Microcanonical averages $\langle h^2\rangle_E$ and $\langle e^2\rangle_E$ as function of the energy per bond for the ferromagnetic ($J=1$) Heisenberg model with $N=1024$ spins on the triangular lattice ($z=6$). This system experiences frustration for $E>0$. The inset shows the parameters $\mu$ and $\eta$ which as in Fig.~\ref{fig:ip_100csl} indicate the positions of phase transitions.}}
\end{center}
\end{figure}

\revis{Finally we applied the method to the Heisenberg model on a triangular lattice with 1024 spins again with 1000 bins. Now the system experiences frustration at positive energies or negative temperatures which for $J=1$ correspond to the antiferromagnet that for this lattice type has a maximal energy $2E_{\rm max}/zN=0.5$. Again the algorithm is able to explore most of the energy range without getting trapped and the time series (not shown here) looks very similar to the previous cases. In Fig.~\ref{fig:tri} the resulting data for $\langle h^2\rangle_E$, $\langle e^2\rangle_E$, and the parameters $\mu$ and $\eta$ are shown.}

\section{Conclusions}

In this study we reviewed how the density of states of a system can be calculated via the inverse microcanonical temperature, i.e., the derivative of the logarithmic density of states, and how the latter can be obtained by means of microcanonical averages. We then introduced an alternative method that avoids mixed derivatives of the Hamiltonian, such that instead of the Hessian only the Laplacian and  the gradient are required thus reducing computational demands. Since the ratio of Laplacian and squared gradient needs to be integrated, preferably with high accuracy, we devised a simple method for numerical integration adapted to the mathematical properties of that function.

Once the density of states can be calculated with sufficient accuracy and precision it can be used to verify the results of established histogram-based methods or -- as shown in this study -- to design a novel flat-distribution Monte Carlo method. This method is similar to the multicanonical method, the Wang-Landau method, or Statistical Temperature MC with the important difference that the information required to bias the ensemble towards a flat distribution is not indirectly obtained through the distribution of energy values but directly measured from the gradient and curvature of the Hamiltonian at the surfaces of constant energy.

The simulations we conducted are intended to be a proof-of-concept and we did not focus on optimizing the algorithm. We deem it likely that improvements can be made in various ways just as there are various histogram-based methods. Even hybrid strategies are conceivable. \revis{We observe that the algorithm is able to produce flat histograms on intervals of energy over which the density of states differs by hundreds to thousands of orders of magnitude, which in turn is convincing evidence that our formula for the density of states is correct and that our method for numerical integration works well for this particular type of function.}

We applied the method first to a system of one hundred interacting Lennard-Jones particles. In order to ensure a stable simulation and converging microcanonical averages we had to exclude the lowest part of the energy spectrum. Nevertheless, even in the current basic form the algorithm was able to cover all three phases -- gaseous, liquid droplet, and frozen crystal-like -- and also managed to  map the low-temperature structural transition of the surface atoms. It turned out that the auxiliary data that are produced during the integration can be used to identity the transitions and the energies at which they occur.

Second we considered the $O(n)$ vector-spin model. After deriving expressions for the Laplacian and gradient of the Hamiltonian it became clear that only the average squared spin energy and the average of the square of a spin's local field are required to calculate the density of states. Both of these can easily be measured during the simulation. 
We conducted a number of simulations for various spin and lattice dimensions and system sizes of up to about a thousand spins. In each case it was easily possible to sample most of the state space. \revis{This was even true for the case of a system with frustration: the Heisenberg model on a triangular lattice.}
We found that the average squared local field depends on $D$ but surprisingly only very little on $n$. 
The defining condition of the microcanonical ensemble is of-course the system's fixed total energy which translates to a known value for the nearest-neighbor spin-spin correlation. This in turn is closely related to the quantities needed to calculate the density of states: The squared local field comprises next-nearest-neighbor spin-spin correlations and the squared local energy the second moment of nearest-neighbor spin products. A more rigorous theoretical analysis of the mutual dependencies of these quantities for the $O(n)$ spin model would be of great interest.

\section*{Acknowledgements}
The project was funded by the Deutsche Forschungsgemeinschaft (DFG, German Research Foundation) through the Collaborative Research Centre under Grant No. 189\,853\,844--SFB/TRR 102 (project B04). S.S. thanks Franziska Facius for hospitality.

\bibliography{text}

\appendix
\section{Calculation of $\langle \MakeLowercase{e}^2\rangle$ for $D=1$}

For $D=1$ and large $N$ the spin products $\pmb{\sigma}_{k-1}\pmb{\sigma}_{k}$ and $\pmb{\sigma}_{k}\pmb{\sigma}_{k+1}$ belonging to adjacent bonds are uncorrelated. The average of one product is given by
\begin{equation}
b_1\coloneqq\langle\pmb{\sigma}_{k}\pmb{\sigma}_{k+1}\rangle=-E/JN.
\end{equation}
To calculate the second moment
\begin{equation}
b_2\coloneqq\langle(\pmb{\sigma}_{k}\pmb{\sigma}_{k+1})^2\rangle
\end{equation}
we consider the partition function
\begin{equation}
Z_n=\int_{-1}^1(1-s^2)^{\frac{n-3}2}e^{as}\diff s,
\end{equation}
with $s=\pmb{\sigma}_{k}\pmb{\sigma}_{k+1}$. One finds
\begin{eqnarray}
Z_2(a) &=& \pi I_0(a),\\
Z_3(a) &=& \frac{2\sinh a}{a},\\
Z_4(a) &=& \frac{\pi I_1(a)}{a},\\
Z_5(a) &=& 4\frac{a\cosh a-\sinh a}{a^3},\\
Z_6(a) &=& \frac{3\pi I_2(a)}{a^2},\\
Z_7(a) &=& 16\frac{(a^2+3)\sinh a-3a\cosh a}{a^5},\\
Z_8(a) &=& \frac{15\pi I_3(a)}{a^3},\\
Z_9(a) &=& 3\frac{32a(a^2+15)\cosh a-(2a^2+5)\sinh a}{a^7},\\
Z_{10}(a) &=& \frac{105\pi I_4(a)}{a^4},
\end{eqnarray}
where $I_k$ are modified Bessel functions. With
\begin{equation}
b_1(a)=\frac{Z'(a)}{Z(a)}
\end{equation}
and
\begin{equation}
b_2(a)=\frac{Z''(a)}{Z(a)}
\end{equation}
and $a\in(-\infty,\infty)$.
One obtains for example with $n=3$
\begin{equation}
b_1(a)=\frac{\cosh a}{\sinh a}-\frac{1}{a}
\end{equation}
and
\begin{equation}
b_2(a)=1-\frac{2\cosh a}{a\sinh a}+\frac{2}{a^2}=1-2\frac{b_1(a)}a.
\end{equation}
The second moment of $e$ is given by
\begin{eqnarray}
\langle e^2 \rangle_E &=& J^2\langle (\pmb{\sigma}_{k-1}\pmb{\sigma}_{k}+\pmb{\sigma}_{k}\pmb{\sigma}_{k+1})^2  \rangle_E,\nonumber\\
                    &=& 2J^2(b_2+b_1^2).
\end{eqnarray}

\end{document}